# Using Orthogonal Channels for Supporting Multicast Service in Multi-channel Wireless Mesh Networks

Abbas Nargesi[1] and Mehdi Ghasemi[2]

[1]Department of IT Engineering, Azad University, Qazvin, Iran
`as.nargesi@gmail.com`
[2] Department of IT Engineering, Azad University, Qazvin, Iran
`mahdi.ghasemi@gmail.com`

**ABSTRACT**

*Unlike wired networks, the capacity of a wireless network is interference limited due to the broadcast nature of wireless medium. Some multicast wireless network protocols do not consider channel assignment issue, that they cause interference at transmission nodes, hence do not use full capacity of the network. Interference can be reduced and throughput improved with the use of multichannel features. Therefore, this paper used orthogonal channels for sending and receiving nodes in the network. We propose EWM (Efficient Wireless Multicast) method that is distributed scheme for constructing multicast tree in multi-channel multi-interface wireless mesh networks (MIMC-WMN) which selects relay nodes and in distributed form assign orthogonal radio channels to them. To more decrease of interference in adding a branch to the tree, the route with minimum end-to-end delay from the source to the multicast receiver will be chosen. Thus, the tree is suitable for multimedia applicants. We also employ the broadcast nature of the wireless media to reduce the number of relay nodes. The proposed algorithm is compared with MCM algorithm in NS2.*

## 1. INTRODUCTION

Wireless networks have become an important technology around the world. This type of networks has widespread applications in the public, military, and business sectors with cheap and reliable products. Similar to the evolution of wired networks, current wireless networks form isolated communication groups without any interconnection among them.Wireless mesh networks (WMN) is recently emerging as an efficient and cost-effective broadband access technology. It is a trustworthy technology in applications such as broadband home networking, network management, and recent transportation systems. There are two types of node in WMN: mesh routers and clients [1]. WMN topology is usually considered as highly stable due to low or no mobility of mesh routers and their unlimited access to power source. Therefore, unlike ad hoc networks which main objectives are minimizing energy consumption and finding an optimal path between mobile nodes, here the main objective is to maximize network throughput. Hence, we need to develop new routing protocols for WMN [1]. Multicast service is devised to efficiently distribute multicast data among group members reducing overall bandwidth utilization in comparison to multi-unicast method. One should note that multicast implementation in wireless environment essentially differs from wired-network mainly due to broadcast nature of the media [2].

Since multicast reduces the number of transmission, it efficiently uses network resource and one should expect lesser delay, higher throughput, and low computation overhead at source and routers. Several mature multicast routing protocols is proposed for multicast support in wired networks. Unfortunately, these protocols cannot be directly employed in WMN. It has been widely accepted that the main design goal in WMN is to maximize throughput [3]. Multicast implementations that assume a single communication channel (or more correctly, protocol that do not consider existence of multiple channels) suffer from collision and interference issues which considerably reduce network throughput and performance. For this reason, several schemes have been proposed for efficient assignment of available channels. It has been proven that channel assignment improves throughput [6]. Work in [4][5] prove that multicast routing in MIMC-WMN is an NP-hard problem and formulate it as an ILP and propose some heuristic scheme for solving it.

MCM [6] is one of the earliest algorithms using dedicated channel assignment strategies to reduce the interference and improve the efficiency of the multicast tree. MCM uses all interface channels when assigning channels to tree links. MCM suffers from two drawbacks namely existence of hidden channels and high processing overhead of channel assignment method. The hidden channel assignment problem is happening since MCM does not consider two-hop neighbours when assigning channels. As a result, if two-hop neighbours are close enough and use a same channel, their signals will interfere. Although MCM tries to solve this problem by proposing a simpler calculating method for channel assignment and transmitting several messages among tree nodes, but its overall processing overhead is still high. In addition, the channel assignment and tree construction algorithms are executed separately.

In our proposal, which is called Efficient Wireless Multicast (EWM), not only channel assignment and tree constructions are performed simultaneously and in distributed manner, but also hidden channel problem is solved for two-hop neighbours without additional cost using a simple approach with low processing overhead. In contrast to similar work [6, 7], one of main objective of EWM is to minimize data delivery delay for multicast data packets. In addition, it utilizes the broadcast nature of wireless media to reduce the number of relay nodes. Our simulations show that EWM performs competently against MCM.

The paper proceeds in section two with advantages of multi-channel multi-interface networks. In section three, we briefly introduce related work. Section four describes the design of EWM. Section five evaluates the EWM considering delay and throughput parameters and section six concludes the paper.

## 2. MIMC-WMN

Recent research reveals that single-interface single-channel WMNs suffer from low throughput due to interference effects [5,6,9]. Sometimes in WMNs there can be interference in a network which has bad impacts on users. To remove this interference, channel diversity has to be used. To obtain channel diversity in a WMN it is highly required to perform channel allocation. Channel allocation can be done by two methods that are fixed channel allocation (FCA) and dynamic channel allocation (DCA). However, these days DCA scheme is preferred. In DCA the call blocking probability is low as compared to FCA [9].
 In a multi-interface WMN, a typical node has multiple interfaces with individual MAC and physical layers. Communications on these interfaces are independent. There are multiple available channels on each interface. Multi-interface facility allows node to have simultaneous transmissions and receptions on multiple its interfaces. And, multi-channels property allows us to reuse used channels (spatial reuse). As a result, network throughput is increased. Such networks are known as multi-channel multi-interface wireless mesh networks (MIMC- WMN).

In contrast to MANET, it is assumed that WMN nodes are stationary. Therefore, the main objective is to maximize throughput of MIMC-WMNs. In order to achieve this goal, one must pay attention to signal interference problem. If two neighbouring nodes use a same channel for transmission, their signal will interfere and a collision occurs which potentially decrements the overall throughput. Also, there are some overlaps between adjacent channels. In fact, there are three orthogonal channels which can be used simultaneously without interference. The design of a multicast protocol that can employ available interfaces and channels efficiently with no or minimum interference is highly desirable [8].

In IEEE 802.11B and G, we say two channels are orthogonal if the difference between their respective numbers is more than five. For example, channels 1, 6, 11 are orthogonal. Setting aside the random method, one of the simplest methods for channel selection and assignment is to partition nodes into three groups and lets members of each group use a same frequency. The main drawback of this approach is that inactive nodes also earn a channel even though they don't use it. We could not assign such a channel to active nodes and as a result throughput of active nodes is decreased [9].

## 3. RELATED WORK

Recent approaches for channel assignment in multicast tree can be classified in three categories. In first category, it is assumed that channels are assigned and the main problem is construction of the multicast tree. There are some methods which assume that the tree exists and we must assign the channels efficiently. Finally, other methods construct the tree and assign the channels all together.

MT3-DA [11] proposes a centralized method which builds the tree using minimum transmission time between adjacent routers. It tries to minimize interference by assigning low interference overlapping channels to routers that have more dependent receivers. In [12], channels are assigned to network nodes before constructing the tree. Then, the algorithm calculates paths considering network topology and performed channel assignments. It also minimizes the number of required transmission. In this method, different channels are assigned to incoming and outgoing links of a relay node.

All the available channels are used in [13] and they consider the number of dependent receivers when selecting relay nodes. In [14], the objective is to minimize interference among multicast nodes and the number of relay nodes. Karimi et al [5] mathematically formulate the problem and solve it using an iterative primal-dual framework based on Lagrange relaxation and primal problem decomposition.

LCA [6], multicast tree is built and then different channels are assigned for node reception and transmission. It uses one channel per each tree level. LCA advantages are ease of implementation and high throughput. But, it suffers from possible interference between member nodes of each level. If the number of channels is more than the number of levels, it wastes extra channels. Then, they propose MCM [6] which selects all available channels so that interference factor is minimized. Since MCM only considers interference of one-hop neighbours, it suffers from hidden channel associated with two-hop neighbours. M4 [7] is proposed to solve hidden neighbour problem of MCM using information of two-hop neighbours. Interference factor of MCM is not a constant value and it is changing over time with transmission rate and network status. In M4, an optimization function is proposed which only use channels number.

## 4. PROPOSED ALGORITHM

Several factors must be taken into account when designing multicast protocols for MIMC-

WMNs. First, we should consider interference effects of indirect neighbours that are simultaneously active when assigning channels. Second, multicast implementation should be fully distributed for scalability reasons. It should also make use of WBA (wireless Broadcast Advantage), interface diversity, and minimize the number of relay nodes. WBA enables a wireless node to broadcast a data packet to all of its direct-neighbours in a single transmission. We use this property to increase the reliability of EWM. Finally as noted earlier, the method should maximize network throughput.

### 4.1. Level Partitioning

We use MCM [6] idea and modify its tree construction and channel assignment methods. The main difference between EWM and MCM is that EWM is a distributed protocol. EWM assumes that all network nodes are potential receivers and run Breadth First Search (BFS) algorithm. As a result, all nodes are portioned in different levels. Then, EWM deletes links between any two nodes in the same level. The algorithm identifies minimal number of relay nodes that form the broadcast tree. In the tree mesh, one node can have more than one parent. The purpose of this step is to identify the only parent (we call it a relay node here) for a node that has more than one parent so that the number of relay nodes is minimal. A top-down approach is used that is from level 0, level 1… to the lowest level to identify the relay nodes. Suppose that we have discovered the relay nodes in level 0, level 1,..., level $i - 1$. Now we describe how to find the relay nodes in level $i$. One can see that fewer relay nodes will result in less traffic flows in the network which means less local interference. Thus, the algorithm is intended to identify the minimal number of relay nodes in level $i$ which can communicate with all the nodes in level $i+1$. The initial network and the network after running level partitioning are shown in Fig.1 and Fig.2 respectively. After running the algorithm, the number of forwarding candidates is reduced in each level (e.g. A and B in level 2). The first two steps of MCM and EWM are the same.

Algorithm 1: Node Search in Level i Algorithm for specify Forward Nodes
**Data**: $T_G(i)$: (i, i+1) sub-tree mesh; $S_i$: nodes in level i; $S_j$: nodes in level i+1
**Result**: R: the set of relay nodes in level i
R= ø;
**While** $S_j$ !=ø do
    In $T_G(i)$, compute the number of parents of each node in $S_j$, and compute
    The number of children of each node in $S_i$;
    Find $v_{i1}, v_{i2},\ldots,$ in $S_j$ with the minimal number of parents;
    Among the parents of $v_{i1}, v_{i2},\ldots,$ find $t_f$ with the maximal number of children;
    R= R ∪ {$t_f$};
    $S_i$= $S_i$ − {$t_f$ };
    The children of $t_f$ record $t_f$ as their relay node;
    $S_j$= $S_j$ − {the children of tf};
**End**

International Journal of Computer Networks & Communications (IJCNC) Vol.4, No.5, September 2012

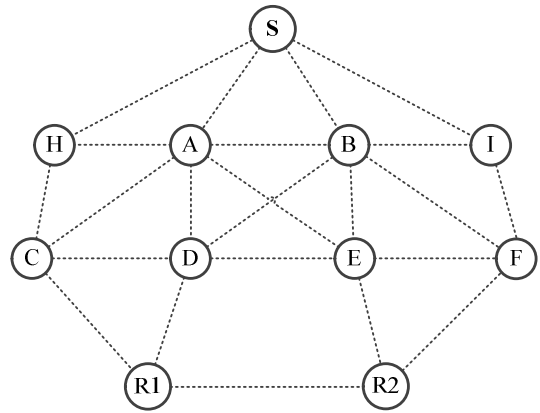

Figure 1. Wireless mesh network

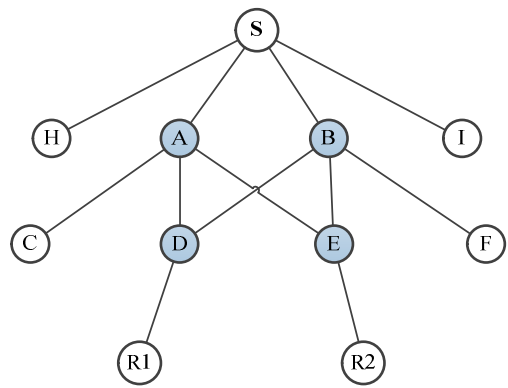

Figure 2. Resulted network after level partitioning

### 4.2. Distributed Tree Construction

After finding candidate relay nodes, the source s must send a periodical ADV message. In each level, candidate relay nodes are responsible to forward the ADV message to nodes of next level. As a result, all nodes receive ADV message. Now, suppose that R1 wants to join multicast group of the source after receiving ADV message. It will send a Join Request (JREQ) message towards its parent. The parent of a node is known when it receives an ADV message. If the parent was a member of the multicast tree, it will respond to the node by sending a Join Response (JREP) message. Otherwise, it forwards the JREQ toward its parent in upper level.

Now, consider a case when some node (such as D) wants to send a JREQ, but it has multiple parents. In another word, the node has received multiple ADV form upper level. In such cases, it will save the address of prospective parents when it receives their respective ADVs. It then selects the sender of first ADV message as parent and sends JREQ toward it. Doing so, the end to end delay from source to each receiver is minimized.

### 4.3. Channel Assignment

There are three orthogonal channels with zero interference in 802.11 standard, namely, 1, 6 and 11. EWM use these channels in order to minimize interference. When source or an intermediate node on the distribution tree sends a JREP in response to a JREQ, it uses next available orthogonal channel for downstream multicast data transmission and reports it via JREP

International Journal of Computer Networks & Communications (IJCNC) Vol.4, No.5, September 2012

message. It means that if our parent uses channel 6, we should use channel 11.

If a node does not receive periodical ADV from source for three consecutive periods, it will assume that the link between it and its parent does not exist anymore. It then inspects next saved ADV messages and selects a new parent. But, if it could not find any other ADV message, it has to send find for next run of EWM level partitioning algorithm to find a new parent. However, it is possible that the new parent use different channel than the previous one. In such cases, the node must update channel assignment of downstream nodes via a proper UPDATE message.

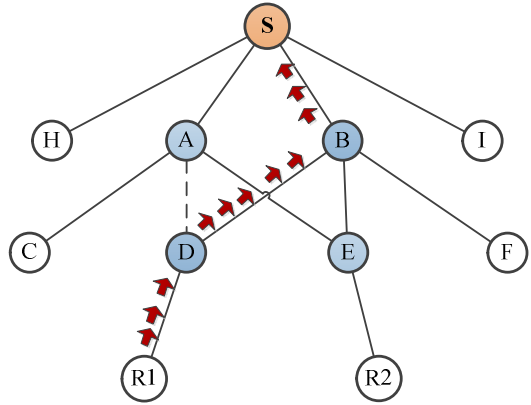

Figure 3. First receiver joins the tree.

### 4.4. An Example

We complete our discussion of EWM with a simple example. In Fig. 3, assume that s sends an ADV. There are four nodes in next level, and nodes A and B should forward the message. This process is repeated until all nodes receive ADV message. Now, R1 joins the group by sending a JREQ toward its parent i.e. D. Since D is not a part of multicast tree, it forwards JREQ to one of its parent. D has two parents i.e. A and B. We assume that D has received ADV message of B sooner. Then, D sends JREQ to B, and finally B sends JREQ to s. This is illustrated in Fig. 3. In response, s selects channel 1, prepares a JREP message and sends it toward B. Then, B selects next available orthogonal channel, i.e. 6, and inserts it in the JREP message. The message is then forwarded to D. The same process is down by D using channel 11 as next available channel. Fig. 4 depicts join process of R1.

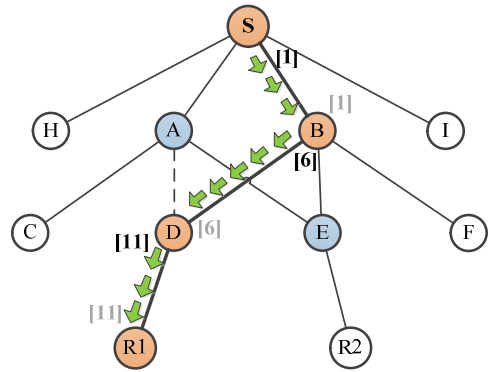

Figure 4. Source responds to JREQ of R1.

Now, if R2 decides to join the group, it sends JREQ to E, and E forwards the message to B. (Fig. 5). Since B is already a member of multicast group, it sends a JREP to E. The response contains the assigned channel number (i.e. 6). As a result, E forwards JREP message announcing its used channel number as 11. The final status of the network is shown in Fig. 6.

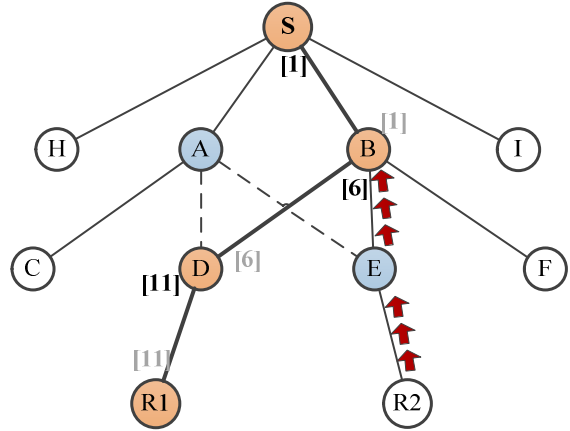

Figure 5. R2 sends a JREQ.

EWM parent selection policy could be modified as follow to minimize the number of relay nodes. First, suppose that all parents of a node are not part of multicast tree. As noted earlier, the node selects a parent among all potential parents whose ADV message is received earlier. But, if some parents were part of the multicast tree, EWM restricts its selection set to those parents. The node can extract this information from received ADV and JREQ messages using WBA. If a node receives both messages from a node, it assumes that the potential parent belongs to the tree. But, if it only received an ADV message and no JREP message, it assumes that this potential parent is not part of the tree. In another word, EWM tries to utilize available relay nodes of the tree whenever possible.

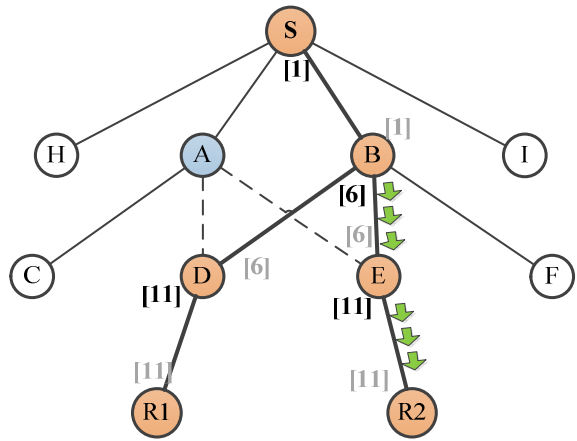

Figure 6. B responds to JREQ of R2.

EWM has several advantages over MCM. First, it solves hidden channel problem associated with two-hop neighbours. Second, it does not rely on centralized manner of computing interference factor and optimization function. MSM first constructs the tree and then assigns channel numbers, both in a distributed manner. Instead, EWM simultaneously constructs the tree and performs channel assignment. Unlike MCM which use all channels, EWM employs only three orthogonal channels. In MCM, there is no available channel for other types of traffic than multicast. EWM assign channels on demand and does not waste channel numbers. We also efficiently use WBA to decrease the number of relay nodes. It means that the number of required transmissions is also minimized. Hence, one should expect lower levels of interference and higher throughput. Finally, MCM does not consider end-to-end delay when constructing the tree.

## 5. SIMULATION RESULTS AND ANALYSIS

In order to evaluate EWM, we have compared it against MCM in MIMC-WMNs using ns-2 simulators. We have used following criteria:

**Throughput**: it is defined as number individual byte received by a receiver divided by time between reception of first packet and last packet. We assume one sender and compute the average of throughput for all receivers.

**End to end delay**: we compute the delivery delay for each packet in each receiver. Then, we compute the average for all packets per all receivers.

**Packet loss**: we compare packet loss rate while increase receiver nodes.

### 5.1. System Model

We model the network as unidirectional graph G (V,T), in which V is set of mesh routers and E is set of edges between neighbour nodes. Each node has two wireless interfaces used for reception and transmission. Number of mesh router is 50 and number of receivers varies between 1 and 45, tab 1 summarize system model.

| PARAMETER | VALUE |
| --- | --- |
| NETWORK SIZE | 50 |
| AREA | 900*900 |
| TRANSMISSION RANGE | 250M |
| GROUP SIZE | 5,10,1,…,45 |
| INTERFACE TYPE | MAC 802.11 |
| SPECTRUM RANGE | 60 MHZ |
| PACKET SIZE | 512 BYTE |
| TRAFFIC MODEL | CBR |

Table1. System model.

## 5.2. End to End delay

Fig. 7 compares end to end delay of EWM against two version of MCM. MCM-n and MCM use n and all channels respectively. EWM delay is lower than MCM-6 and similar to MCM. But, EWM only uses 3 channels instead of 12. It is shown that using more channels is very effective in reducing end to end delay [6,7,8], since it directly reduces interference. But, EWM achieves same delay as MCM with much less channels. It is because EWM considers end to end delay when constructing the tree. Each downstream node saves received ADV messages in time order and selects the owner of first one as its parent. Since all nodes use this criterion, on should expect that the delay of selected path between source and receiver is minimum.

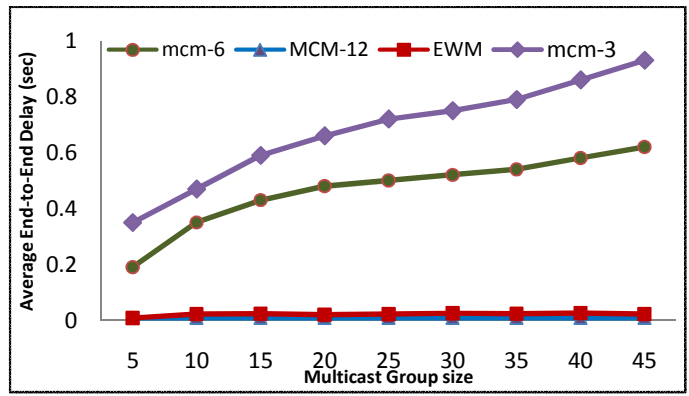

Figure 7. Comparison of end to end delay for various group sizes.

## 5.3. Throughput

It is known that using multi-channel and multi-interface nodes increases throughput. In Fig. 8, clearly all methods shows good throughput. Also, throughput decreases with number of receivers as expected. EWM throughput is better than MCM-3 since it avoids hidden channels of two-hop neighbours. EWM also efficiently use WBA property of network. It also assigns channels to those nodes that need them. Essentially, MCM is appropriate when number of used channels is high. However, MCM outperforms EWM, because it uses all interface channels.

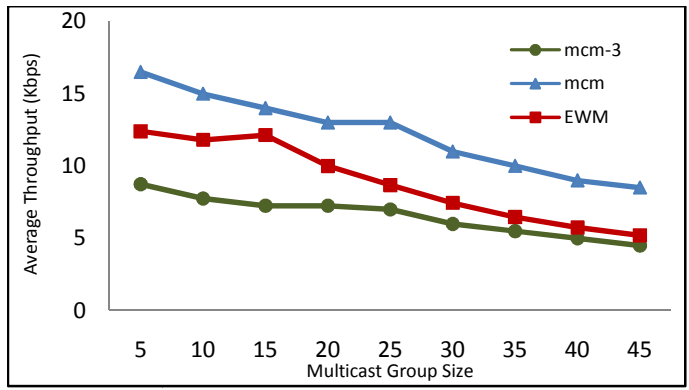

Figure 8. Effect of group size on throughput.

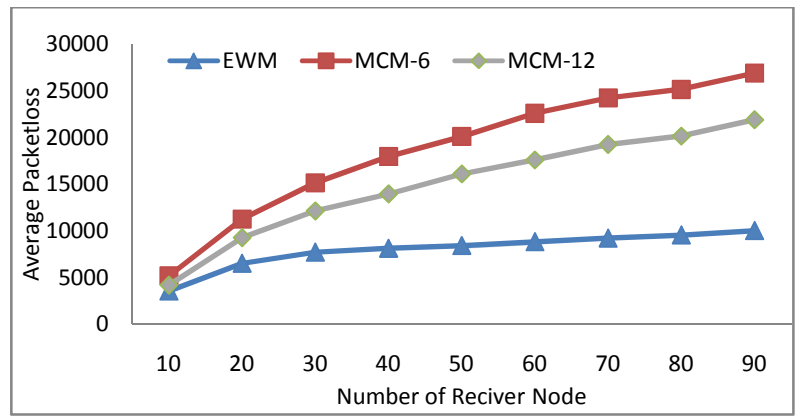

Figure 9. Effect of group size on packet loss.

The proposed algorithm solves the HCP problem and hidden channel problem in MCM algorithm which can cause collision in sending packets. Fig. 9 compares packet loss of EWM against two version of MCM. In fig. 9 when the number of receiver nodes increases, the packet loss rate is almost linear for the proposed algorithm but when the number of receiver and forwarding nodes increases, hidden channels are too high and as a result the packet loss rate increases in the 6-channel or 12-channel MCM.

## 6. CONCLUSION

WMN technology proved to be a revolutionary and modern technology which has remarkable impacts on telecommunication and internet systems. Channel assignment is a new important topic in WMN because multi-channel multi-interface approach increases network capacity. We proposed a distributed multicast protocol for MCMI-WMN which exhibits a low delivery delay and high throughput making it a good candidate for supporting multimedia application. EWM uses a limited number of available channels leaving others free for other traffic types while achieving acceptable throughput. Our algorithm solves the MCM problem like HCP, end to end delay and free channel for other traffic in the network. Moreover, MCM first constructs a tree and then performs the channel assignment while the proposed algorithm does these steps altogether. Our simulation shows that EWM is a viable alternative for MCM when the number of available channels is limited.

## REFRENCES